\documentclass[twocolumn,twocolappendix]{aastex63}
\usepackage{amsmath}
\usepackage{amssymb}
\usepackage{bm}
\usepackage{wasysym}
\usepackage{hyperref}
\hypersetup{colorlinks,breaklinks,allcolors=blue}
\usepackage[super]{nth}
\usepackage{wrapfig}
\usepackage{graphicx}
\usepackage{grffile}
\usepackage{longtable}
\usepackage{multirow}
\usepackage{xspace}

\newcommand{\pnas}{Proc. Nat. Aca. Sci. USA}
\newcommand{\psj}{Planet. Sci. J}

\newcommand{\swift}{{\sc Swift}\xspace}
\newcommand{\woma}{{\sc WoMa}\xspace}
\newcommand{\seagen}{{\sc SEAGen}\xspace}

\newcommand{\moon}{{\scriptsize\leftmoon{}}}

\shorttitle{Immediate origin of the Moon}
\shortauthors{Kegerreis et al.}

\begin{document}

\title{\Large Immediate origin of the Moon as a post-impact satellite}

\correspondingauthor{Jacob Kegerreis}
\email{jacob.kegerreis@durham.ac.uk}
\author[0000-0001-5383-236X]{J. A. Kegerreis}
\affiliation{Physics Department, Institute for Computational Cosmology, Durham University, Durham, DH1 3LE, UK}
\affiliation{NASA Ames Research Center, Moffett Field, CA, USA}

\author[0000-0003-0925-9804]{S. Ruiz-Bonilla}
\affiliation{Physics Department, Institute for Computational Cosmology, Durham University, Durham, DH1 3LE, UK}
\author[0000-0001-5416-8675]{V. R. Eke}
\affiliation{Physics Department, Institute for Computational Cosmology, Durham University, Durham, DH1 3LE, UK}
\author[0000-0002-6085-3780]{R. J. Massey}
\affiliation{Physics Department, Institute for Computational Cosmology, Durham University, Durham, DH1 3LE, UK}
\author[0000-0002-4630-1840]{T. D. Sandnes}
\affiliation{Physics Department, Institute for Computational Cosmology, Durham University, Durham, DH1 3LE, UK}
\author[0000-0002-8346-0138]{L. F. A. Teodoro}
\affiliation{BAERI/NASA Ames Research Center, Moffett Field, CA, USA}
\affiliation{School of Physics and Astronomy, University of Glasgow, G12 8QQ, Scotland, UK}




\begin{abstract}

The Moon is traditionally thought to have coalesced from the debris ejected by a giant impact onto the early Earth. However, such models struggle to explain the similar isotopic compositions of Earth and lunar rocks at the same time as the system's angular momentum, and the details of potential impact scenarios are hotly debated. Above a high resolution threshold for simulations, we find that giant impacts can immediately place a satellite with similar mass and iron content to the Moon into orbit far outside the Earth's Roche limit. Even satellites that initially pass within the Roche limit can reliably and predictably survive, by being partially stripped then torqued onto wider, stable orbits. Furthermore, the outer layers of these directly formed satellites are molten over cooler interiors and are composed of around 60\% proto-Earth material. This could alleviate the tension between the Moon's Earth-like isotopic composition and the different signature expected for the impactor. Immediate formation opens up new options for the Moon's early orbit and evolution, including the possibility of a highly tilted orbit to explain the lunar inclination, and offers a simpler, single-stage scenario for the origin of the Moon.

\end{abstract}

\keywords{
  Lunar origin(966);
  Impact phenomena(779);
  Earth--Moon system(436);
  Hydrodynamical simulations (767).
}


\section{Introduction} \label{sec:introduction}

In the canonical hypothesis for the origin of the Moon,
the early Earth is hit by a Mars-sized impactor, `Theia'
\citep{Hartmann+Davis1975,Cameron+Ward1976,Canup+2021}.
The collision ejects a debris disk
that can explain the Moon's large mass, angular momentum, and tiny iron core;
but it creates a Moon derived mostly from impactor material
\citep{Canup+Asphaug2001,Canup+2021}.
This is a concern because
the Moon has a near-identical isotopic composition to the Earth
for many elements \citep{Melosh2014,Meier+2014,Lock+2020},
and it seems unlikely, though perhaps possible,
that the impactor would already match the proto-Earth target's composition
\citep{Dauphas2017,Mastrobuono-Battisti+Perets2017,Schiller+2018,Johansen+2021}.
Additional equilibration after the impact could help,
but is probably insufficient
\citep{Pahlevan+Stevenson2007,Nakajima+Stevenson2015}.
However, some recent analysis suggests distinctly different oxygen isotopes
with increasing depth in the lunar mantle \citep{Cano+2020},
and hydrogen isotopes also indicate imperfect mixing
between the proto-Earth and Theia \citep{Desch+Robinson2019}.

The relative difference between the compositions
of the resulting Moon (or proto-lunar disk) and Earth is often expressed by
$\delta f_{\rm t} \equiv (f_{\rm t}^\moon / f_{\rm t}^\oplus) - 1$,
where $f_{\rm t}^{\moon\!,\oplus}$ is the mass fraction
of each silicate reservoir that originated in the target \citep{Reufer+2012},
such that $|\delta f_{\rm t}| < 10$\% indicates very similar compositions
and a pure-Theia lunar mantle would have $\delta f_{\rm t} = -100$\%.

Alternative impact scenarios have been proposed to improve
results above the $\delta f_{\rm t} \approx -70$\% of canonical models.
High angular momentum impacts
into rapidly spinning targets \citep{Cuk+Stewart2012,Lock+2018}
can eject and mix more proto-Earth material,
as can a very large impactor \citep{Canup2012}.
The excess angular momentum might be removable
in or near the evection resonance,
but removing the correct amount may be difficult \citep{Rufu+Canup2020}.
Hit-and-run impacts can also make somewhat more target-rich disks
\citep{Reufer+2012}.
Multiple impacts could create successive intermediate satellites
that combine to form the Moon \citep{Rufu+2017},
depending on the merger efficiency \citep{Citron+2018}.
Somewhat separately,
a proto-Earth magma ocean could be more readily injected into orbit
\citep{Hosono+2019},
and numerical effects might be inhibiting mixing in simulations
\citep{Deng+2019a}.
The circumstances required for some of these events
may also have low likelihoods,
though with only one Moon to study
we are reminded that its origin could be improbable \citep{Melosh2014}.

Numerical simulations of giant impacts commonly use
smoothed particle hydrodynamics (SPH)
to model planets using particles
that evolve under gravity and pressure.
Most previous Moon-formation simulations
have used around $10^5$--$10^6$ particles,
but these resolutions can fail to converge
on even large-scale outcomes of giant impacts,
such as the planet's rotation period or the mass of ejected debris
\citep{Genda+2015,Hosono+2017,Kegerreis+2019}.
Here we use up to $10^8$ particles.
At this resolution, each particle has a mass of $6 \times 10^{16}$~kg
and an effective size in the planet of $\sim$14~km.
A lunar-mass satellite itself would be composed of around $10^6$ particles,
which enables us to inspect its composition in detail.

Directly produced satellites were found
in some early simulations of giant impacts \citep{Benz+1987,Canup+Asphaug2001},
but have typically been dismissed in terms of lunar formation
\citep{Canup+Asphaug2001,Asphaug2014} because of
(1) then-justified low-resolution numerical concerns;
(2) a lack of iron and of proto-Earth material;
(3) overly fine-tuned requirements for the impact parameters;
(4) and/or orbits that crossed interior to the Roche limit.
In contrast, here with orders of magnitude more particles,
we find that stable satellites are produced
(1) reliably at high numerical resolutions of at least $10^7$ SPH particles;
(2) with a Moon-like mass of iron and significant proto-Earth material;
(3) over a small but appreciable range of impact angles and speeds
especially for spinning planets \citep{RuizBonilla+2021};
(4) and that satellites with Roche-interior initial trajectories
often survive partial disruption to be torqued onto wider final orbits.

\newpage
\section{Methods} \label{sec:methods}

\subsection{Initial conditions}

The proto-Earth target and Theia impactor are differentiated into
an iron core and a rocky mantle
containing 30\% and 70\% of the mass, respectively,
with fiducial canonical-like masses of 0.877 and 0.133~$M_\oplus$.
We use the updated ANEOS Fe$_{85}$Si$_{15}$ (Earth-core analogue) and forsterite
equations of state (EoS) \citep{Stewart+2019},
which encompass thermodynamically consistent (no-tension) models
of multiple phases and improved fits to experimental data.
The default temperature at the surface is set to 2000~K,
for a mantle specific entropy of 2.9~kJ~K$^{-1}$~kg$^{-1}$.
The internal profiles are adiabatic
with continuous temperature across the core--mantle boundary.

The planets' internal profiles are generated by integrating inwards
while maintaining hydrostatic equilibrium%
\footnote{
  The \woma code \citep{RuizBonilla+2021}
  for producing spherical and spinning planetary profiles and
  initial conditions is publicly available with documentation and examples at
  \href{https://github.com/srbonilla/WoMa}{github.com/srbonilla/WoMa},
  and the python module \texttt{woma} can be installed directly with
  \href{https://pypi.org/project/woma/}{pip}.
},
then the roughly equal-mass particles
are placed to precisely match the resulting density profiles
using the stretched equal-area method%
\footnote{
  \seagen \citep{Kegerreis+2019} is publicly available at
  \href{https://github.com/jkeger/seagen}{github.com/jkeger/seagen},
  or as part of \woma.
},
and its modified version for spinning planets \citep{RuizBonilla+2021}.
Before simulating the impact, a brief 10~ks settling simulation is first run
for each body in isolation, to allow any final settling to occur.
The specific entropies of the particles are kept fixed,
enforcing that the particles relax themselves adiabatically.

\subsection{Smoothed particle hydrodynamics (SPH) simulations}

We run the $\sim$400 impact simulations in this study using the
open-source hydrodynamics and gravity code \swift%
\footnote{
  \swift \citep{Schaller+2018} is publicly available at
  \href{www.swiftsim.com}{www.swiftsim.com}.
}.
Most of these use a `vanilla' form of SPH
plus the \citet{Balsara1995} switch for the artificial viscosity
\citep{Kegerreis+2019},
to a simulation time of 90, 120, or 180~ks (25--50~hours)
depending on whether e.g. tidal stripping
or secondary impact events have concluded,
in a cubic box of side 120~$R_\oplus$.
Any particles that leave the box are removed from the simulation.

To test the sensitivity of our results to the numerical methods
and to known challenges for SPH,
an additional set of comparison simulations are run using:
(1) the boundary-improvement methods of \citet{RuizBonilla+2022},
where a simple statistic is used to identify and then correct particles
with inappropriate SPH densities near material interfaces and free surfaces,
which mitigates the problems raised by density discontinuities in standard SPH;
and (2) the geometric density average force (GDF) expression
for the SPH equations of motion \citep{Wadsley+2017},
which can further improve behaviour near sharp density gradients.

\begin{figure*}[t]
  \centering
  \includegraphics[
  width=\textwidth, trim={92mm 32.2mm 70mm 7mm}, clip]{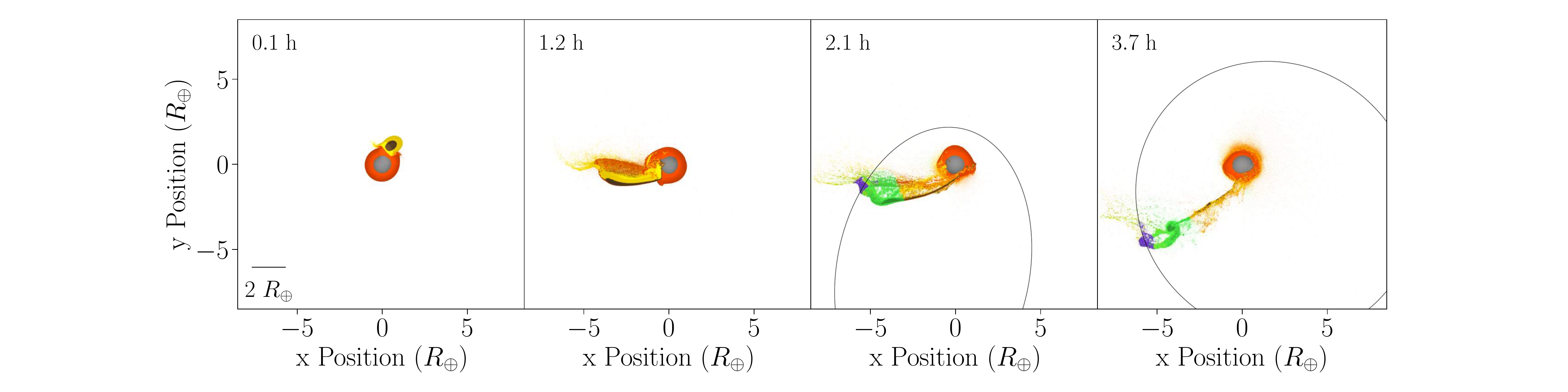}\\\vspace{-0.5mm}
  \includegraphics[
  width=\textwidth, trim={92mm 32.2mm 70mm 7.5mm}, clip]{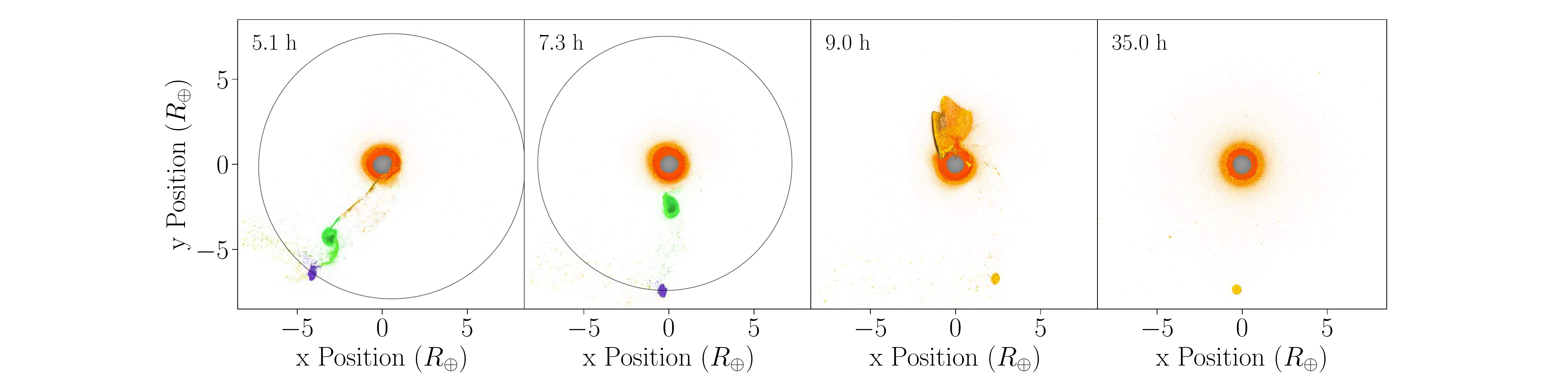}
  \\\vspace{-0.5em}
  \caption{
  Illustrative snapshots from an impact simulation
  where a satellite is placed directly onto a wide orbit,
  in this example with the lowest final eccentricity.
  In the middle panels, the particles that will form the satellite
  and inner remnant are highlighted in purple and green.
  The black lines show the estimated orbit.
  Grey and orange show the proto-Earth's
  core and mantle material respectively,
  and brown and yellow the same for Theia.
  The colour luminosity varies slightly with the internal energy.
  The annotated time is measured from first contact,
  so the simulation began at $-1$~h.
  An animation is available at
  \href{http://icc.dur.ac.uk/giant_impacts/moon_wide_orbit_slice.mp4}{icc.dur.ac.uk/giant\_impacts/moon\_wide\_orbit\_slice.mp4},
  and with the same data rendered in 3D at
  \href{http://icc.dur.ac.uk/giant_impacts/moon_wide_orbit_houdini.mp4}{icc.dur.ac.uk/giant\_impacts/moon\_wide\_orbit\_houdini.mp4}.
  \label{fig:B46v102_snaps_etc}}
\end{figure*}

\subsection{Impact scenarios and simulation suites}

We start from a base scenario similar to a canonical Moon-forming impact:
an impact angle of $\beta=45^\circ$;
a speed at contact of the mutual escape speed,
${v_{\rm c} = 1~v_{\rm esc}}$ ($\sim$$9$~km~s$^{-1}$);
masses for the target proto-Earth and impactor Theia
of $M_{\rm t} = 0.887$ and $M_{\rm i} = 0.133$~$M_\oplus$;
and no pre-impact spins,
as illustrated in Fig.~\ref{fig:impact_scenario}.

The largest set of scenarios finely covers a focused range of angles and speeds:
$\beta = 43, 44, 45, 46, 47, 48^\circ$ and
$v_{\rm c} = 0.98, 1.00, 1.02, 1.04$~$v_{\rm esc}$.
This corresponds to angular momenta from $1.19$--$1.37$
times that of the present-day Earth--Moon system,
$L_{\rm EM} = 3.5 \times 10^{34}$~kg~m$^2$~s$^{-1}$.
To examine the numerical effects
of our finite-particle model planets,
each scenario in this set is repeated 8 times
using a rotated orientation for the settled target or impactor.
For the other sets of simulations,
rather than attempting a complete sampling
of the many-dimensional parameter space,
we explore a coarser range of more highly varied values,
to test whether our conclusions are sensitive
to significant changes of the body masses, spins, and temperatures.

We run a similar angle exploration for different-mass bodies,
using $\tfrac{3}{4}$ and $\tfrac{1}{2}$ of the base Theia's mass
and keeping the total mass constant:
$M_{\rm i} = 0.100$ and $0.067~M_\oplus$,
and corresponding $M_{\rm t} = 0.920$ and $0.953~M_\oplus$.
For these smaller impactors,
qualitatively similar outcomes arise at larger impact angles.
As such, we test
$\beta = 45$--$50^\circ$ for the $\tfrac{3}{4}$ mass Theia,
and $\beta = 52$--$57^\circ$ for the $\tfrac{1}{2}$ mass Theia,
in $1^\circ$ increments.

We then return to the base scenario
and vary the pre-impact spin of each body,
for spin angular momenta of
$L_{\rm t, i} = -\tfrac{1}{2}, -\tfrac{1}{4}, \tfrac{1}{4},
\tfrac{1}{2}$~$L^{\rm max}_{\rm t, i}$,
where $L^{\rm max}$ corresponds to the maximum stable spin
(minimum period) for these planets \citep{RuizBonilla+2021}:
$1.0 \times 10^{35}$~kg~m$^2$~s$^{-1}$ (2.3~h) for the proto-Earth,
and $4.9 \times 10^{33}$~kg~m$^2$~s$^{-1}$ (2.5~h) for Theia.
The $\tfrac{1}{4}$ and $\tfrac{1}{2}$~$L^{\rm max}$ spin periods
are 5.2 and 3.0~h for the proto-Earth and 5.4 and 3.2~h for Theia.
The spinning bodies' axes of rotation are set in the same direction as the
orbital angular momentum of the impact ($+z$),
such that the impact point is on the equator
and the spin is expected to have the greatest overall effect on the outcome.
One additional proof-of-concept simulation with $10^{7.5}$ particles
is run with the proto-Earth's spin angular momentum
($\sim$$\tfrac{1}{4}~L^{\rm max}_{\rm t}$) 
in the $+y$ direction, misaligned from the orbit
such that Theia collides with primarily north-pole material.

We also test different temperature profiles for both planets,
repeating the first set of angle and speed scenarios
using surface temperatures of 1000, the base 2000~K, and 3000~K,
yielding profiles below, slightly below, and above solidus, respectively.
These lower and higher temperatures
increase and decrease the bodies' average densities
by about 3 and 7\% for the proto-Earth,
and 4 and 10\% for Theia,
as well as testing the sensitivity
to different regimes in the equation of state.

Finally, we repeat a subset of the first suite of scenarios
to study numerical reliability and convergence.
The sets described above all use $\sim$$10^7$ SPH particles,
compared with the $10^5$--$10^6$ typically used in the current literature,
which can be insufficient to resolve or converge on
both small and large-scale outcomes of giant impacts
\citep{Genda+2015,Hosono+2017,Kegerreis+2019}.
We repeat the $\beta = 44$--$47^\circ$,
$v_{\rm c} = 1.00, 1.02$~$v_{\rm esc}$ scenarios
with $10^{4}$, $10^{4.5}$, $10^{5}$, $10^{5.5}$, $10^{6}$, $10^{6.5}$,
$10^{7.5}$, and $10^{8}$ particles.
The same scenarios are also repeated for $10^7$ particles using
the modified SPH scheme with the GDF \citep{Wadsley+2017}
and boundary-improvement modifications \citep{RuizBonilla+2022}.
In addition, for the base scenario for each resolution below $10^{7.5}$
we also run 8 reoriented repeats,
as done for the full $10^7$-particle angle and speed suite.

\begin{figure*}[t]
	\centering
  \vspace{-0.7em}
	\includegraphics[
    width=\textwidth, trim={191mm 32mm 191mm 7mm}, clip]{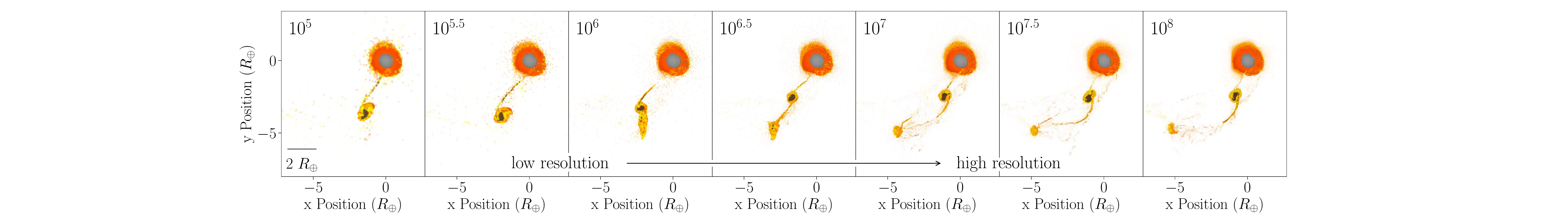}
  \\\vspace{-0.5em}
  \caption{
    The categorically distinct behaviour of outer-satellite formation
    emerges consistently for resolutions
    above a threshold of $>$$10^{6.5}$ particles.
    Each panel shows a snapshot
    from simulations of the same scenario
    at the same time 3.6~h after impact
    using different numbers of particles.
    $10^{4}$ and $10^{4.5}$ particles (not shown) behave similarly to $10^5$.
		\label{fig:B45v100_res_snaps}}
\end{figure*}

\section{Results and discussion} \label{sec:results}

\subsection{Immediate satellite formation}

We find that a key feature of impact scenarios that launch a large satellite
directly into a wide orbit
is the early separation of the proto-satellite
from the main remnant of the impactor.
This behaviour emerges reliably with sufficient numerical resolution.
The inner remnant then transfers angular momentum to the satellite
of ejected proto-Earth and Theia material
and slingshots it into orbit,
as illustrated in Fig.~\ref{fig:B46v102_snaps_etc},
before falling back to re-impact the target.

The initial satellite separation
is consistent for simulations with over $10^{6.5}$ SPH particles,
up to and including our highest resolution of $10^8$,
as shown in Fig.~\ref{fig:B45v100_res_snaps}.
A full investigation of what determines
the exact number of particles required to resolve adequately
the tidal and hydrodynamic evolution
at this scale is left for future work.
Lower resolutions instead produce a single larger remnant
that stays intact until it grazes or re-impacts the proto-Earth
to produce a spray of debris.

In the particular scenario of Fig.~\ref{fig:B46v102_snaps_etc},
the satellite actually overtakes the inner remnant briefly
around 5~h (fifth panel),
and the now-reversed torque slightly shrinks and circularises the final orbit.
This resulting satellite has a mass of $0.69~M_\moon$
and a nearly circular orbit with a periapsis of $7.1~R_\oplus$,
far outside the Roche limit of $\sim$$2.9~R_\oplus$
and even likely beyond the evection and eviction resonances
for a moderate rotation period \citep{Touma+Wisdom1998,Cuk+Stewart2012}.
The observed transfer of angular momentum is also matched well by
simple estimates for the forces between the orbiting bodies
(Appendix~\ref{sec:appx:torques}, Fig.~\ref{fig:B46v102_orbit_evol}).

This direct formation of a satellite is sensitive to the impact angle,
with milder dependencies on the speed and initial spins,
as detailed further in Appendix~\ref{sec:appx:explore}.
Large satellites are best produced at impact angles around $45^\circ$
-- the most likely angle for a generic impact --
and near the mutual escape speed.
For non-spinning planets,
the range of satellite-producing impact angles is only a few degrees,
but for spinning planets, wider ranges become feasible.
In particular, a prograde spinning proto-Earth
allows less disruption of Theia for viable satellite formation at lower angles,
although high initial rotation rates may require
some angular momentum to be removed from the final system
\citep{Rufu+Canup2020}.

The initial temperature and internal structure of the proto-Earth and Theia
do not substantially affect these results.
Even a much smaller Theia with $\tfrac{3}{4}$ the mass
still produces similar stable satellites, at larger impact angles.
Similar outcomes were also found using different equations of state
\citep{RuizBonilla+2021},
and using a modified version of SPH that mitigates known issues
at boundaries between materials and vacuum surfaces \citep{RuizBonilla+2022}.
The region of parameter space for the immediate formation
of stable satellites is not huge,
but appears to be numerically robust
and not restricted to any low-likelihood parameter values.

\begin{figure}[t]
	\centering
	\includegraphics[
    width=\columnwidth, trim={8mm 7.5mm 7.5mm 6mm}, clip]{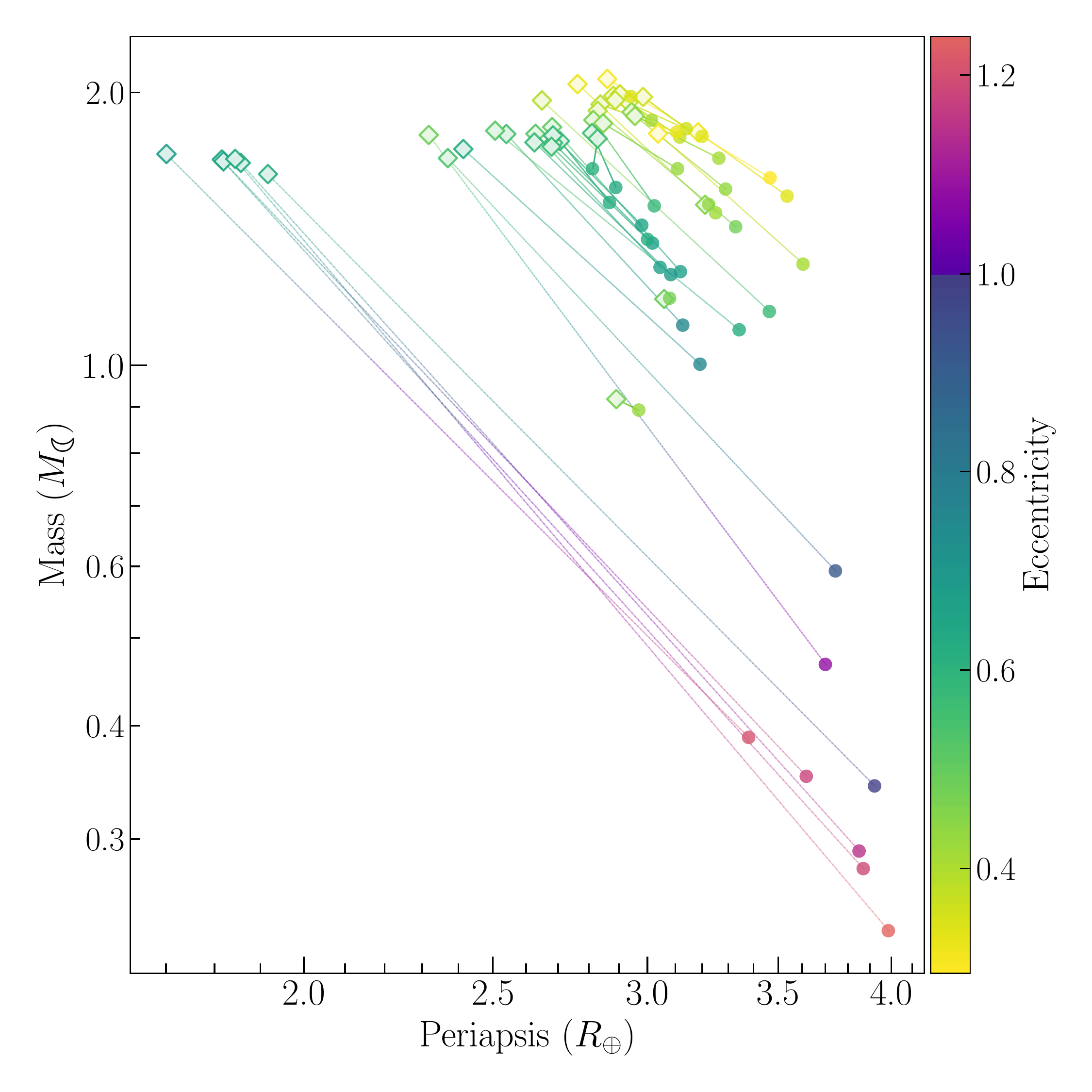}
  \\\vspace{-0.5em}
  \caption{
    The change in mass, periapsis, and orbital eccentricity
    of satellites that pass through a periapsis near or inside the Roche limit
    and retain at least 10\% of their initial mass,
    which tends to result in a significantly wider final orbit
    following partial tidal stripping.
    Diamonds and circles show the pre- and post-periapsis satellites,
    respectively.
		\label{fig:periapsis_passings_m_r_p_e}}
  \vspace{-1em}
\end{figure}

\begin{figure}[t]
  \centering
  \includegraphics[
  width=\columnwidth, trim={8mm 8mm 7.5mm 7mm}, clip]{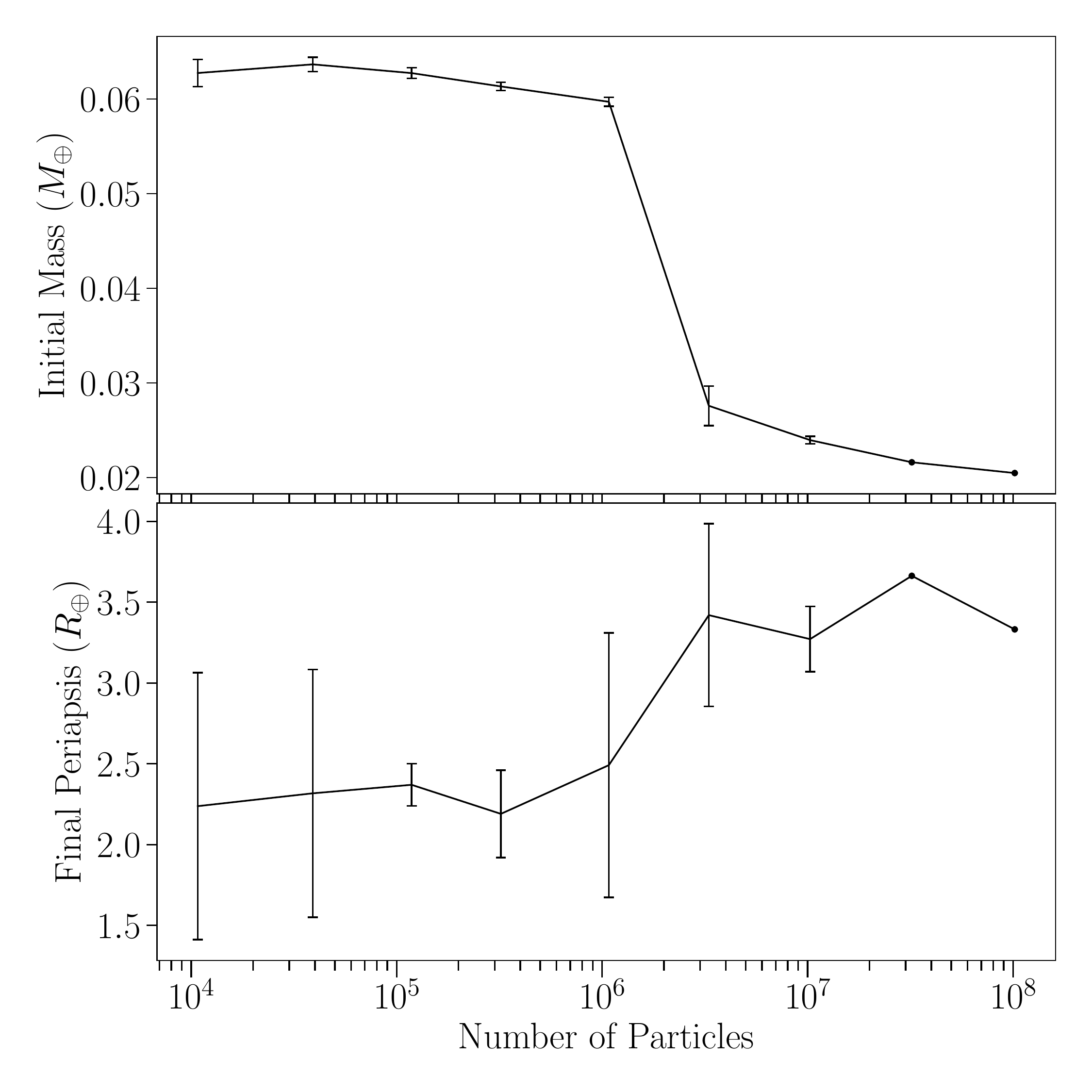}
  \\\vspace{-0.5em}
  \caption{
    The convergence and uncertainty of
    the mass of the initial satellite (or larger single remnant)
    and the periapsis of the final satellite
    that form in different-resolution simulations
    of the impact scenario in Fig.~\ref{fig:B45v100_res_snaps},
    as a function of the number of particles.
    The error bars show the standard deviations across eight reoriented repeats,
    with points for the individual $10^{7.5,\,8}$-particle simulations.
  \label{fig:B45v100_res_trends}}
  \vspace{-1em}
\end{figure}

\subsection{Tidal stripping onto stable orbits}

In some scenarios, the satellite is not launched out quite as far,
and its periapsis falls within the Roche limit.
However, not only can these satellites survive partial tidal disruption
on their initial orbit,
but the stripped material
can transfer angular momentum to the surviving satellite
and torque it onto a stable, Roche-exterior final orbit.
This significantly extends the parameter-space range of scenarios
that produce a Moon-like satellite.

The change in mass, periapsis, and eccentricity
from all large satellites that pass near or inside the Roche limit
and survive with at least 10\% of their initial mass
is shown in Fig.~\ref{fig:periapsis_passings_m_r_p_e}.
Depending on their initial orbits,
satellites may lose little-to-no mass;
lose some mass and be torqued onto stable orbits,
from initial periapses as low as $\sim$$2.4~R_\oplus$;
or suffer near-total disruption and be ejected onto unbound orbits.
The estimated Roche limit of $2.9~R_\oplus$,
which assumes a circular orbit,
is a decent prediction for the furthest distance
at which these satellites on eccentric orbits
begin to be partially disrupted.

In most cases, the partial stripping and resulting transfer of angular momentum
mildly increase the eccentricity of the final satellite.
However, a significant minority have their eccentricity reduced,
usually when the stripping occurs late enough
that a torque is still applied as the satellite approaches apoapsis.
Regardless, the periapsis distance is raised
in every case of significant mass loss,
often to well beyond the Roche limit.

This general behaviour is remarkably consistent given the diversity
in mass, initial orbit, spin, and debris environment
of the pre-periapsis satellites,
with a dominant dependency on the initial periapsis, $r_{\rm p}$.
We find that the fraction of mass that survives
passage inside the Roche limit, $m_{\rm f}/m_{\rm i}$,
is fit well by a simple analytical prediction
(Fig.~\ref{fig:periapsis_passings_dm_r_p_spin}),
derived in Appendix~\ref{sec:appx:torques:predictions}:
\begin{equation}
  \dfrac{m_{\rm f}}{m_{\rm i}} \approx
    \dfrac{r_{\rm p}^3}{t_{\rm Roche}} \sqrt{\dfrac{2 \pi \rho}{3 G M^2}} \;,
  \label{eqn:periapsis_dm}
\end{equation}
where $t_{\rm Roche}$ is the time spent within the Roche limit
as predicted from the initial orbit,
$\rho$ the satellite's density,
and $M$ the planet's mass.

Across the tested impact scenarios,
the major outcomes and conclusions for forming stable satellites
are consistent for simulations with $\geq$$10^7$ particles.
However, these multi-stage collisions and stripping events are somewhat chaotic;
small changes to the initial satellite
can have larger consequences for its evolution.
We probe this uncertainty
by running additional simulations of the same scenario
with reoriented initial conditions,
which with infinite resolution should give identical outcomes.
Higher resolutions do reduce the scatter in results,
but the precise values for the masses and orbits of the final satellites
are not yet perfectly converged,
as shown in Fig.~\ref{fig:B45v100_res_trends}.
At standard/low resolutions below $10^{6.5}$ particles,
the formation of the initial satellite is not resolved reliably,
so instead a wider variety of low-mass final bodies are formed.
At higher resolutions, some initially similar satellites
may dip within the Roche limit briefly enough
to suffer minimal tidal stripping,
while others with just slightly lower periapses
can have enough material removed to be torqued onto somewhat wider orbits
(Fig.~\ref{fig:reoriented_orbits}).

The longer-term evolution of these satellites will largely depend on
the tidal interactions and angular momentum transfer
from the rapidly spinning Earth and the disk
\citep{Goldreich+Tremaine1980,Touma+Wisdom1994},
and perhaps on friction with extended disks on shorter timescales.
Additionally, larger disks might produce other moonlets
\citep{Rufu+2017,Citron+2018b},
and satellites may accrete significant additional material from the debris disk
\citep{Salmon+Canup2012,Lock+2018}.

\subsection{Satellite compositions and interiors}

Canonical Moon-forming scenarios produce debris disks
composed of only $\sim$30\% proto-Earth material \citep{Canup+2021},
which is difficult to reconcile with the near-identical isotopic signatures
of the Earth and Moon.
Here, the immediate satellites typically have
moderately higher bulk compositions of around 30--40\% proto-Earth material.
Furthermore, most show a strong gradient in provenance with radius,
with a deep interior of mostly Theia material
under a roughly linearly increasing and isotropic proportion
of proto-Earth mantle
\citep{RuizBonilla+2021}.
The remaining debris disks (see Table~\ref{tab:results}),
from which a satellite may later accrete more mass,
contain similar amounts of proto-Earth and Theia material.

In otherwise Moon-like satellites,
the outer $\sim$10\% by radius can reach over 60\% proto-Earth material.
The outer 30\% by radius ($\sim$$\tfrac{2}{3}$ the mass)
then averages to around 40--50\% proto-Earth.
This corresponds to a depth of about 500~km,
relatable to the $\sim$300--1000~km depths that are considered
for the lunar magma ocean \citep{Charlier+2018,Johnson+2021},
within which an initial gradient may be expected to mix.

The outer material is molten, typically heated to at least 4000~K by the impact,
but the deeper interior is only a few hundred Kelvin
above its initial near-solidus state,
in contrast with the fully molten satellites
expected to accrete from most debris disks
\citep{Pritchard+Stevenson2000,Lock+2020}.
Our cooler, default, and warmer planet initial conditions
thus produce satellites with mostly solid,
sub-liquidus melt, and molten interiors, respectively.

The deep interiors of all our simulated satellites
contain some iron from the impactor's core.
For satellites with masses similar to the Moon,
the typical iron content ranges from around 0.1 to 3\%,
comparable with the $\sim$1\% mass of the lunar core \citep{Williams+2014}.

\begin{figure}[t]
	\centering
	\includegraphics[
    width=\columnwidth, trim={8mm 7.5mm 7.5mm 6mm}, clip]{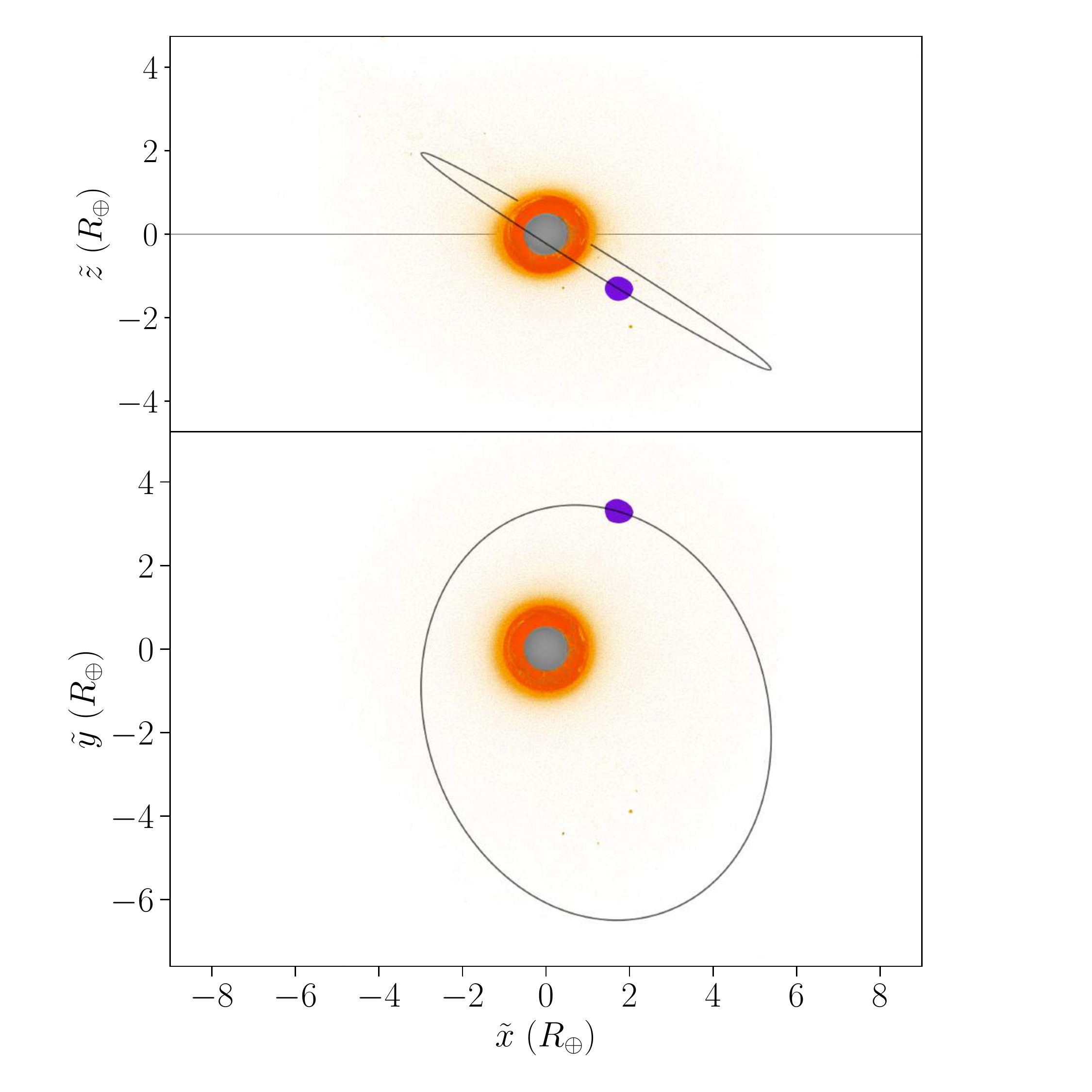}
  \\\vspace{-0.3em}
  \caption{
    The inclined satellite produced by an impact onto a spinning proto-Earth
    misaligned to Theia's orbital angular momentum,
    shown as in Fig.~\ref{fig:B46v102_snaps_etc}.
    The $\tilde{z}$ direction is set as
    the orientation of the post-impact planet's spin angular momentum,
    which is estimated simply as the total
    over all particles within $2~R_\oplus$.
    The results are not sensitive to this choice,
    as using $1.5$ or $2.5~R_\oplus$
    only changes the inferred satellite inclination by $<0.5^\circ$.
		\label{fig:L075y_snap}}
  \vspace{-1em}
\end{figure}

\subsection{Satellite inclination from misaligned target spin}
Previous Moon-formation scenarios yield debris disks
that are closely aligned with the Earth's equator.
This conflicts with the initial orbital tilt away from the equator
of $10^\circ$ or higher that could explain
the Moon's present-day $\sim$$5^\circ$ inclination 
\citep{Touma+Wisdom1998,Cuk+2016a}.
This issue has prompted alternative suggestions involving
resonances with debris or the Sun \citep{Ward+Canup2000,Tian+Wisdom2020},
or close encounters with planetesimals \citep{Pahlevan+Morbidelli2015}.

In contrast, we find that an impact onto a spinning target
with angular momentum misaligned to that of Theia's orbit
can readily produce significantly inclined debris including a satellite,
as illustrated in Fig.~\ref{fig:L075y_snap}.
In this example, the satellite is created similarly
to the equatorial cases described above,
with a mass of $1.4~M_\moon$, a periapsis of $3.1~R_\oplus$,
and an inclination of $32^\circ$.

\section{Implications and Conclusions} \label{sec:conclusions}
To determine whether these satellites
can explain other properties of the Moon
in addition to the mass and iron content,
such as those without fully-molten interiors,
bespoke future studies are required
to extrapolate the simulation outputs reliably to the present day
-- as remains an ongoing challenge
for standard debris-accretion models as well \citep{Lock+2020,Canup+2021}.
With this in mind, here we briefly speculate on the possible implications
and distinctions from other scenarios.

If the gradient of proto-Earth material in the satellite
is entirely mixed away,
then we would find $\delta f_{\rm t} \approx -60$\%,
only slightly better than canonical models.
However, if it does survive even partially,
depending on some imperfect extent of radial mixing
that may be helped by the cooler interior,
then this improves $\delta f_{\rm t}$ to around $-40$ to $-30$\%,
similar to proposed hit-and-run scenarios \citep{Reufer+2012}.
If the outer $\sim$$0.15~R_\oplus$ of Earth's mantle also remains distinct,
which could help to explain geochemical heterogeneities
\citep{Nakajima+Stevenson2015,Deng+2019b},
then our $\delta f_{\rm t}$ would increase by an additional $\sim$$10$\%.
A `hit-and-run-return' collision can both give a higher likelihood
for a canonical-speed final impact like those considered here,
and raise $\delta f_{\rm t}$ by another $\sim$$10$\% \citep{Asphaug+2021},
yielding $\delta f_{\rm t} \approx -20$ to $-10$\%
for these immediate satellites.
Magma oceans on the proto-Earth may raise this even further \citep{Hosono+2019}.
This result could resolve the isotopic conundrum
for a range of Theia compositions
even if only a subset of these processes are effective
\citep{Meier+2014,Asphaug+2021},
but warrants in-depth study of the long-term thermal and tidal evolution.

These directly formed satellites also provide a hitherto overlooked
range of initial conditions for the Moon's early evolution:
with a wider and/or more eccentric or inclined orbit
-- outside the Roche radius
and potentially also the evection and eviction resonances --
and the option of a solid or partial-melt interior.
For example, the Moon's thin crust
may not be consistent with the fully molten Moon
expected from the accretion of hot debris in other models
\citep{Pritchard+Stevenson2000,Charlier+2018,Johnson+2021}.
A cohesive interior and non-circular orbit
might also help to explain the lunar fossil figure,
depending on the extent of tidal heating \citep{Matsuyama+2021}.
In addition, the compositional gradient
aligns with measurements of less Earth-like isotopes
in the deep lunar mantle \citep{Cano+2020}.
The lunar volatile signature may be difficult to reproduce
without a prolonged disk phase \citep{Dauphas+2022},
but loss from a magma ocean might be sufficient \citep{Day+2020},
and significant disk material could also be later accreted
onto the satellite's sampled exterior \citep{Salmon+Canup2012,Citron+2018b}.
Finally, a satellite on a wide, significantly inclined orbit,
which we demonstrate can be produced by a misaligned pre-impact spin,
could preserve its inclination to help to explain the Moon's tilted orbit
\citep{Cuk+2016a,Tian+Wisdom2020}.

In conclusion, high-resolution simulations reveal how
giant impacts can immediately place a satellite into a wide orbit
with a Moon-like mass and iron content.
The resulting outer layers rich in proto-Earth material
and the new options opened up for the initial lunar orbit and internal structure
could help to explain the isotopic composition of the Moon
and other unsolved or debated lunar mysteries.
The system's angular momentum can range from the present-day to higher values,
especially as the spin of the proto-Earth is increased.
Satellites that pass inside the Roche limit
can predictably survive on new, higher-periapsis orbits.
This extends the range of scenarios that can produce Moon-like satellites,
and is a relevant process to consider in other planetary systems.
The likelihood and potential of this and other Moon-formation scenarios
will be constrained by:
more reliable models for the long-term evolution of
satellite orbits, magma oceans, post-impact planets, and disks;
further improved equations of state
in high-resolution simulations across more of the wide parameter space;
and deeper understanding of the isotopic and other constraints
from existing and future measurements \citep{Lock+2020,Canup+2021}.

\acknowledgments

This work was supported by a DiRAC Director’s Discretionary Time award,
by Science and Technology Facilities Council (STFC)
grants ST/P000541/1 and ST/T000244/1,
and used the DiRAC@Durham facility managed by the
Institute for Computational Cosmology
on behalf of the STFC DiRAC HPC Facility (www.dirac.ac.uk).
This equipment was funded by BEIS via STFC capital grants
ST/K00042X/1, ST/P002293/1, ST/R002371/1 and ST/S002502/1,
Durham University and STFC operations grant ST/R000832/1.
DiRAC is part of the National e-Infrastructure.
We also thank Kevin Zahnle for comments,
and Paul Silcox for VFX advice.
The research in this paper made use of the \swift open-source simulation code
\citep{Schaller+2018}, version 0.9.0.
J.A.K. acknowledges support from STFC grants ST/N001494/1 and ST/T002565/1
and a NASA Postdoctoral Program Fellowship.
S.R-B. is supported by STFC grant ST/P006744/1 and Durham University.
V.R.E. and R.J.M. are supported by STFC grant ST/T000244/1.
T.D.S. is supported by STFC grants ST/T506047/1 and ST/V506643/1.
L.F.A.T. acknowledges support from
NASA Emerging Worlds Program award 80NSSC18K0499.

%



\software{
  \swift (\href{www.swiftsim.com}{www.swiftsim.com},
  \citet{Schaller+2018}, version 0.9.0);
  \woma (\href{https://pypi.org/project/woma/}{pypi.org/project/woma/},
  \citet{RuizBonilla+2021}).
}



\newpage
\appendix

\begin{figure}[t]
	\centering
	\includegraphics[
    width=0.95\columnwidth, trim={3cm 8cm 2.5cm 6.5cm}, clip]{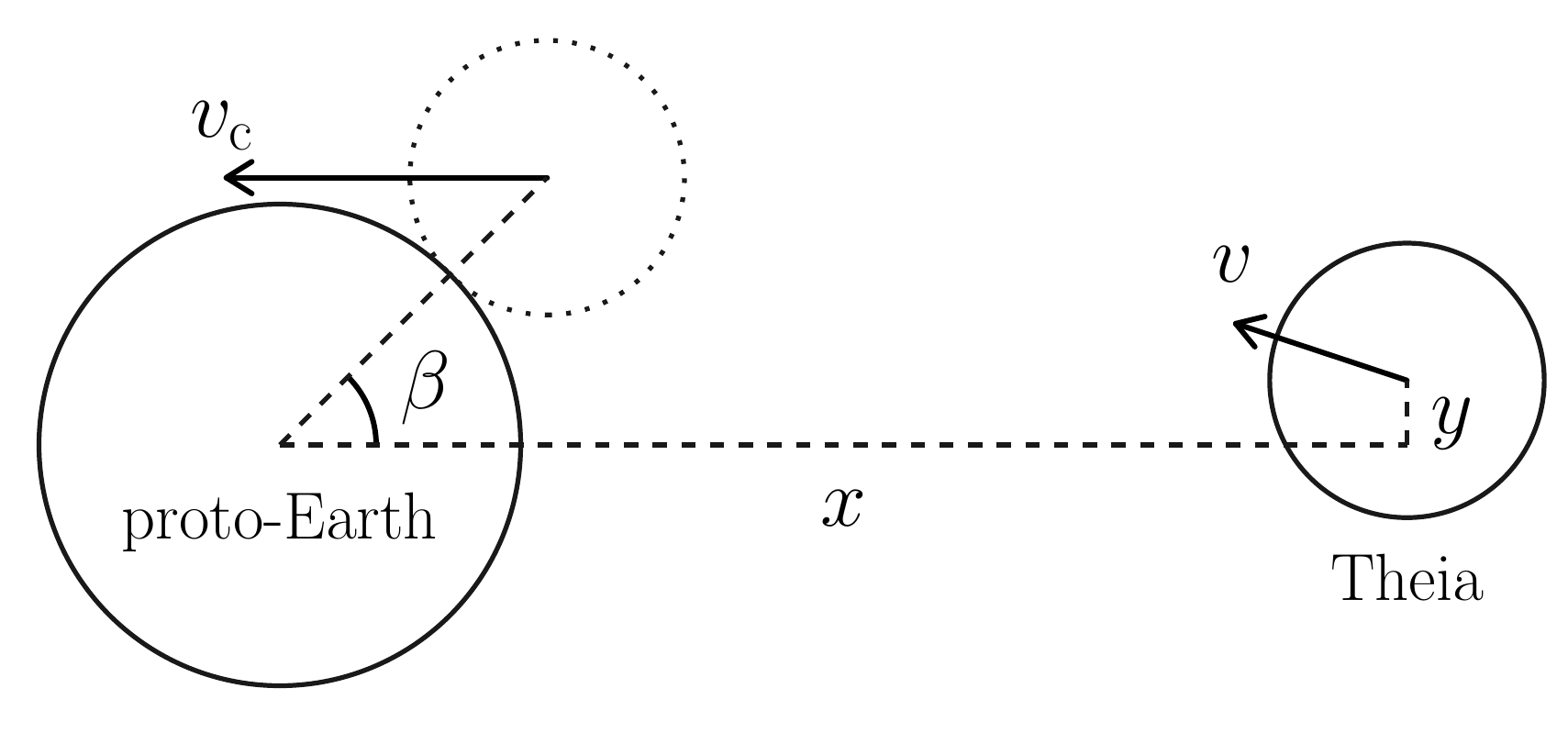}
	\caption{The initial conditions for an impact scenario, to scale,
    in the proto-Earth target's rest frame.
    The angle and speed at first contact, $\beta$ and $v_{\rm c}$,
    are set neglecting any tidal distortion before the collision.
    The initial separation is set such that the time to impact is 1~hour,
    as detailed in \citet[][Appendix~A]{Kegerreis+2020},
    with the velocity at contact in the $-x$ direction.
    \label{fig:impact_scenario}}
\end{figure}

Fig.~\ref{fig:impact_scenario} shows a diagram of the impact initial conditions.

\counterwithin{figure}{section}

\section{Torques and tidal stripping} \label{sec:appx:torques}

\begin{figure}[t]
	\centering
	\includegraphics[
    width=\columnwidth, trim={73mm 16mm 30mm 35mm}, clip]{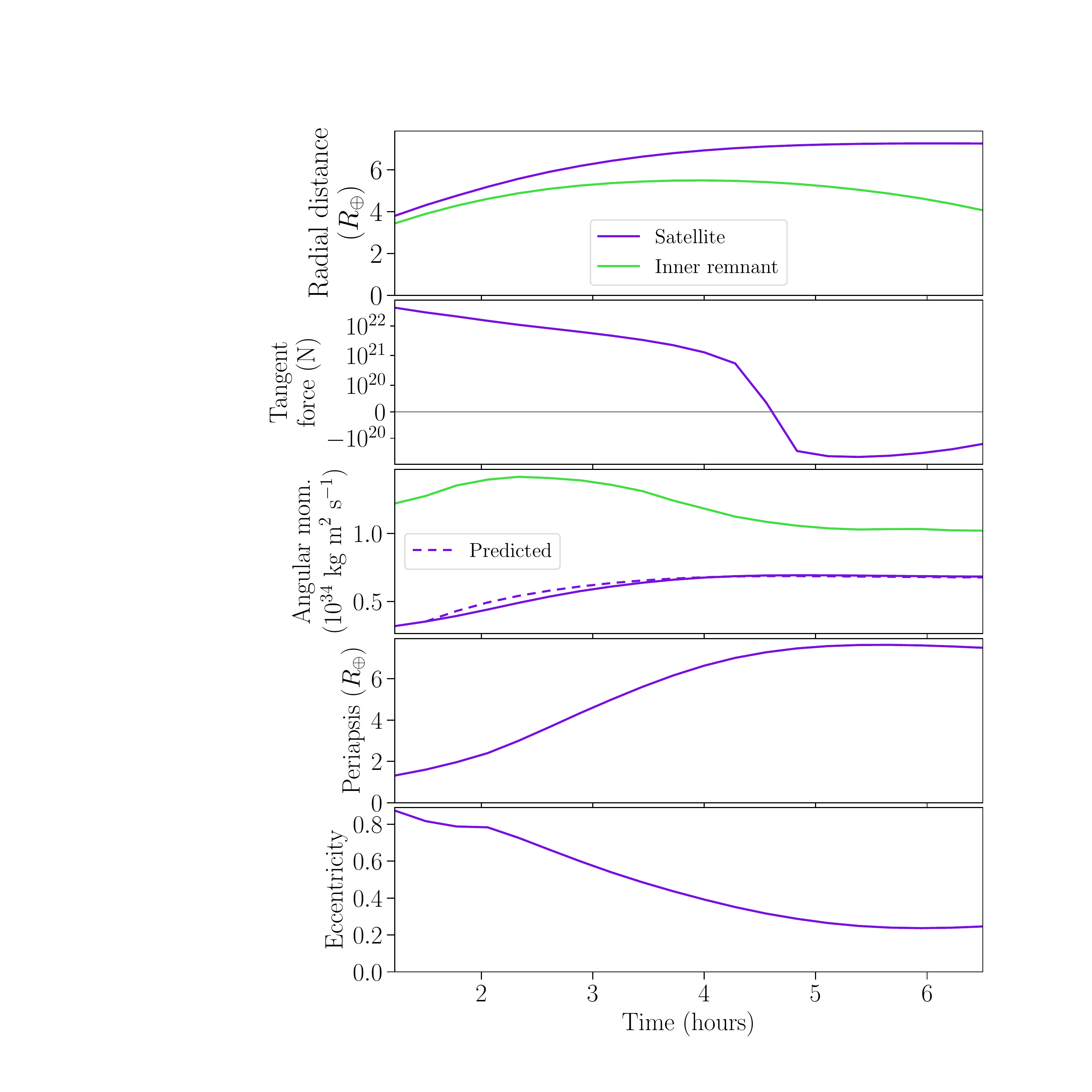}
  \caption{
    The early orbital evolution of the forming satellite and inner remnant
    highlighted in matching colours in Fig.~\ref{fig:B46v102_snaps_etc}.
    The panels show: the radial distances of the bodies from the central planet;
    the tangential component of the force from the inner body on the satellite;
    the change in each body's angular momentum
    and the prediction for that of the satellite from the tangential force
    (dashed line);
    the satellite's periapsis;
    and its eccentricity.
		\label{fig:B46v102_orbit_evol}}
\end{figure}

The main text describes how a satellite on a wide orbit is produced
in scenarios like the one illustrated by Fig.~\ref{fig:B46v102_snaps_etc}.
The orbital evolution of the interacting outer satellite and inner remnant
in that example are shown in Fig.~\ref{fig:B46v102_orbit_evol}.
To estimate the evolution of the satellite's orbital angular momentum,
we integrate the acceleration from the tangential component of the force
from the inner remnant,
i.e. the force perpendicular to the radial vector to the planet.
The predicted increase of the satellite's orbital angular momentum
matches the simulation fairly well
(dashed line in Fig.~\ref{fig:B46v102_orbit_evol}).
This estimate uses the centres of mass taken from the already-run simulation,
so the agreement is more of an encouraging confirmation
that the torque explains the evolution
than a predictive model for generic scenarios.

The satellite and inner-remnant particles are selected
using a standard friends-of-friends (FoF) algorithm
with a linking length of $0.017~R_\oplus$.
The selected masses and results are not affected by small changes to this value.
For final satellites,
we require a minimum of 50 particles to form a FoF group.

\begin{figure*}[t]
	\centering
	\includegraphics[
    width=\textwidth, trim={98.6mm 29.2mm 65mm 5mm}, clip]{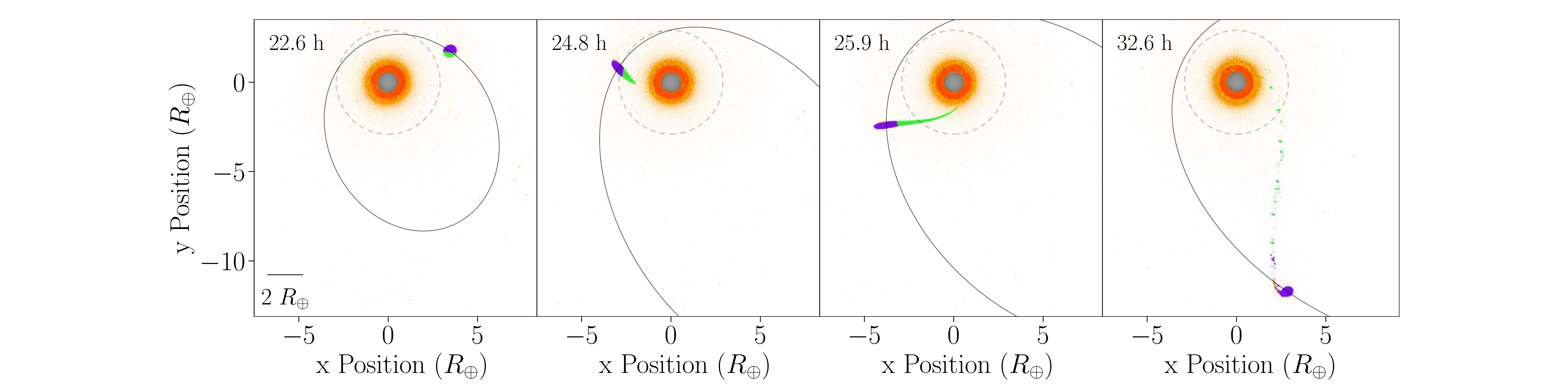}
  \caption{
    Illustrative snapshots of the first periapsis passage after the
    satellite-forming inital impact,
    from a simulation where the final satellite ends up on a stable orbit.
    Colours and annotations are the same as in Fig.~\ref{fig:B46v102_snaps_etc},
    except here purple and green highlight the material
    that will end up in the final satellite
    and that will be stripped from the initial body, respectively.
    The dashed line indicates the Roche limit.
    An animation is available at
    \href{http://icc.dur.ac.uk/giant_impacts/moon_strip_orbit_slice.mp4}{icc.dur.ac.uk/giant\_impacts/moon\_strip\_orbit\_slice.mp4},
    and with the same data rendered in 3D at
    \href{http://icc.dur.ac.uk/giant_impacts/moon_strip_orbit_houdini.mp4}{icc.dur.ac.uk/giant\_impacts/moon\_strip\_orbit\_houdini.mp4}.
		\label{fig:B45v102_strip_snaps}}
\end{figure*}

\begin{figure}[h!]
	\centering
	\includegraphics[
    width=\columnwidth, trim={73mm 16mm 30mm 35mm}, clip]{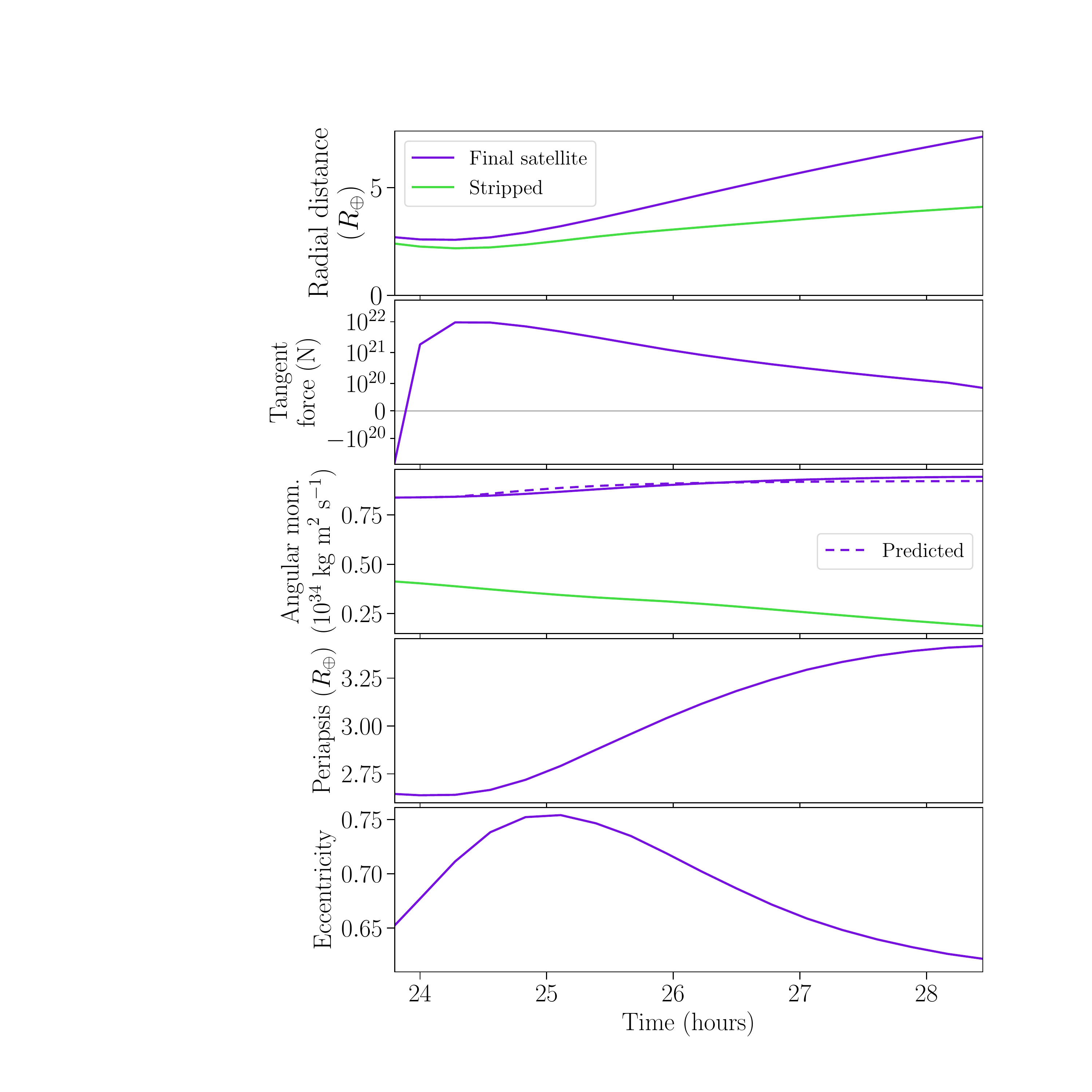}
  \caption{
    The orbital evolution of the material
    that survived and that was stripped from the initial satellite
    in the periapsis passage illustrated in Fig.~\ref{fig:B45v102_strip_snaps},
    presented as in Fig.~\ref{fig:B46v102_orbit_evol}.
		\label{fig:B45v102_orbit_evol}}
\end{figure}

\subsection{Tidal stripping onto a stable orbit}

If the torque from the inner remnant is not quite enough
to raise the satellite's orbit outside the Roche limit of $\sim$$2.9~R_\oplus$,
then as it passes through periapsis it will be tidally disrupted,
to some extent.
However, the stripped material
can exchange angular momentum with the surviving body
and torque it again onto a more stable orbit,
as illustrated in Fig.~\ref{fig:B45v102_strip_snaps}.
In this example, the initial satellite
falls deep inside the Roche limit
with a periapsis of $2.6~R_\oplus$.
It loses about one third of its mass, stripped out into a long tidal tail,
for a final mass of $1.1~M_\moon$.
The surviving satellite continues on a new orbit
with its periapsis raised outside the Roche limit to $3.1~R_\oplus$,
protecting it from further disruption.

As with the previous example,
we can examine this evolution using the centres of mass
of the two pieces, shown in Fig.~\ref{fig:B45v102_orbit_evol}.
Here, the predicted angular-momentum evolution
is still a helpful sanity check
but matches the simulation a bit less precisely,
perhaps because a point mass is a worse approximation
for the elongated tail of stripped material
than a more condensed inner remnant.

\begin{figure}[t]
	\centering
	\includegraphics[
    width=\columnwidth, trim={36.8mm 32mm 5mm 12mm}, clip]{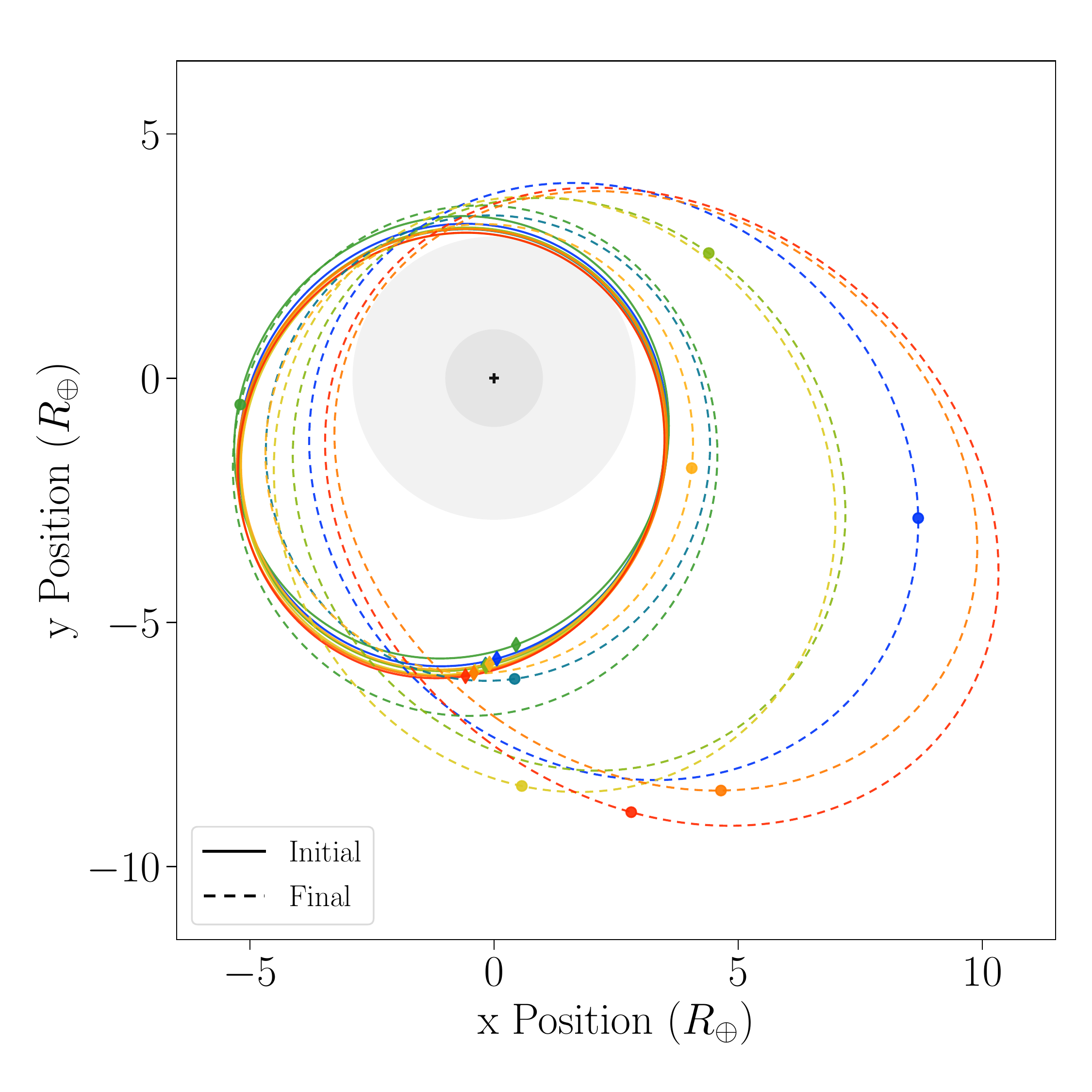}
  \caption{
    The positions and orbits of the satellites
    produced by eight reoriented repeats of the same scenario
    as in Figs.~\ref{fig:B45v100_res_snaps} and \ref{fig:B45v100_res_trends}
    with $10^7$ particles,
    at the same times
    before (solid lines and diamonds) and after (dashed lines and circles)
    the first periapsis tidal-stripping event.
    The inner and outer grey circles indicate the planet's radius
    and the Roche limit at about $1$ and $2.9~R_\oplus$, respectively.
		\label{fig:reoriented_orbits}}
\end{figure}

This type of tidal stripping process exacerbates the differences
in the detailed outcomes between reoriented repeat simulations,
as discussed with respect to Fig.~\ref{fig:B45v100_res_trends} in the main text.
The orbits of satellites before and after periapsis
from the same scenario are shown in Fig.~\ref{fig:reoriented_orbits},
illustrating how similar initial satellites
can separate into a wider spread of final results
-- although the primary conclusions
for the creation of a large stable satellite remain consistent.

\subsection{Tidal stripping trends and predictions} \label{sec:appx:torques:predictions}

As mentioned in the main text,
we find a wide variety of satellites that
suffer different amounts of tidal disruption by falling within the Roche limit
and may be torqued onto wider orbits
(see Fig.~\ref{fig:periapsis_passings_m_r_p_e}).
These pre-periapsis satellites produced naturally from the initial impacts
do not evenly tile the parameter space,
but they cover enough of a range for us
to examine the general behaviour and trends,
and to derive and test a simple model for the fraction of lost mass.

Note also that this empirical set includes scenarios with, for example:
significantly different-size and different-composition bodies;
other debris also in orbit;
and rapidly spinning initial satellites;
all of which likely add noise to these results.
A more thorough and systematic exploration is left for future work.

\begin{figure}[t]
	\centering
	\includegraphics[
    width=0.9\columnwidth, trim={1cm 7.3cm 1cm 8cm}, clip]{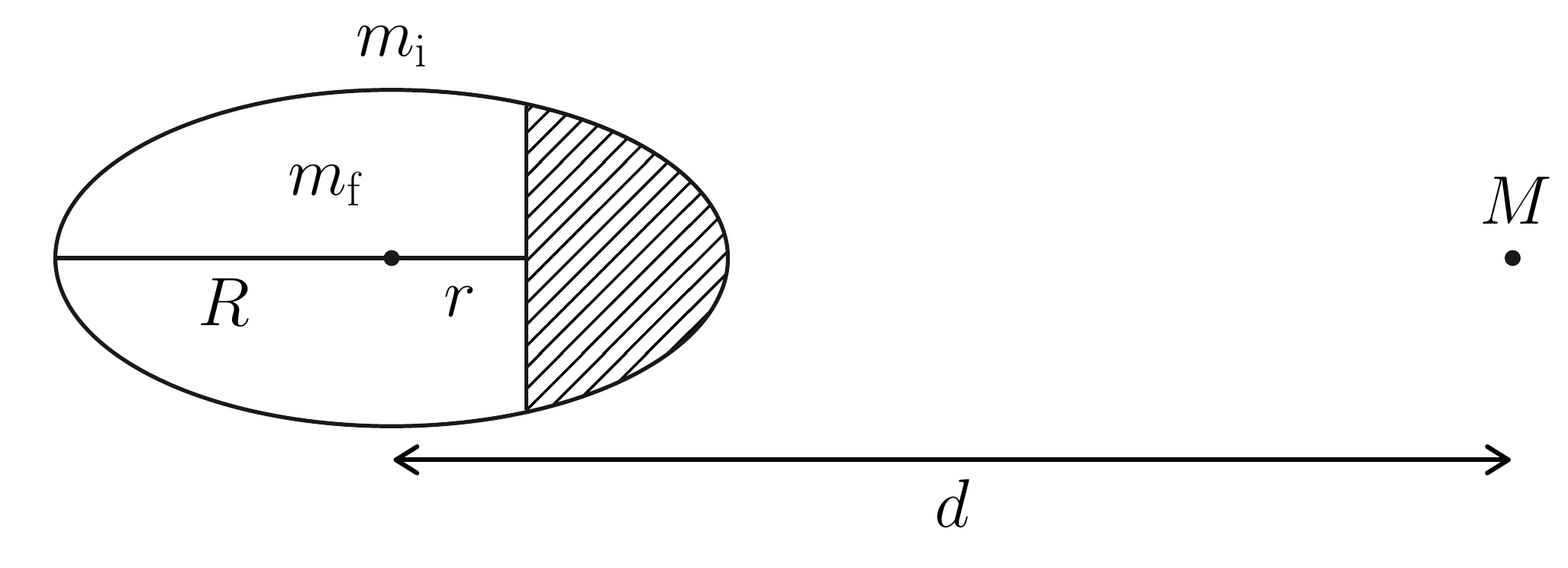}\\
	\includegraphics[
    width=0.9\columnwidth, trim={0.8cm 7cm 4cm 4.5cm}, clip]{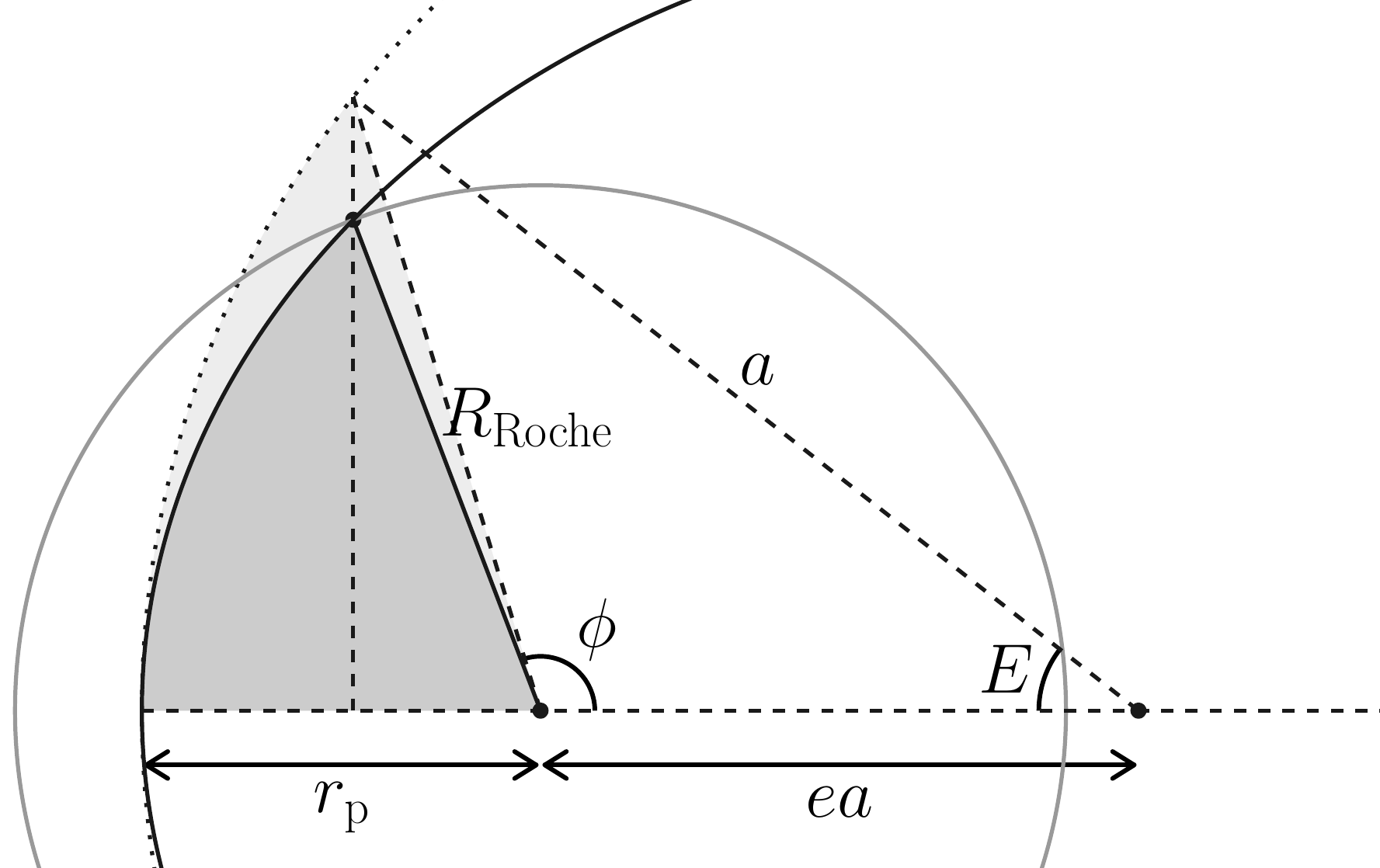}
  \caption{
    Diagrams for estimating the stripping of material
    when a satellite passes through periapsis within the Roche limit.
    (top) Notation for the mass lost (hatched region) from the
    tidal force at a distance $r$ from the centre of a satellite with radius $R$
    and initial mass $m_{\rm i}$,
    at a distance $d$ from a planet with mass $M$.
    (bottom) Notation for the time spent inside the Roche limit,
    $R_{\rm Roche}$ (grey circle),
    by a satellite on an orbit (black ellipse) with a semi-major axis $a$,
    eccentricity $e$, and a periapsis $r_{\rm p} < R_{\rm Roche}$.
    $\phi$ is the complement to the true anomaly
    and $E$ is the eccentric anomaly.
    The dotted line shows the auxiliary circle.
		\label{fig:periapsis_diagrams}}
\end{figure}

At a distance $d$ from the planet,
at a point $r$ from the centre of the satellite
-- see the diagrams in Fig.~\ref{fig:periapsis_diagrams} --
the tidal acceleration is $G M 2 r / d^3$.
Crudely approximating this acceleration as constant for the time the satellite
spends inside the Roche limit, $t_{\rm Roche}$,
and setting the distance equal to the periapsis
to calculate the maximum acceleration, $d = r_{\rm p}$,
we assume that the material will be removed if it is accelerated to the
escape speed of the satellite, $\sqrt{2 G m_{\rm i} / R}$,
where $m_{\rm i}$ and $R$ are its inital mass and radius.
We further approximate the final mass as proportional to the
radius beyond which material is lost, $m_{\rm f}/m_{\rm i} \approx r / R$,
equivalent to treating the satellite as a cylinder that is cleanly split.
This yields
\begin{equation}
  \dfrac{m_{\rm f}}{m_{\rm i}} \approx
    \dfrac{r_{\rm p}^3}{t_{\rm Roche}} \sqrt{\dfrac{2 \pi \rho}{3 G M^2}} \;.
\end{equation}
The time spent inside the Roche limit can be estimated
using the area of the shaded region
in the bottom panel of Fig~\ref{fig:periapsis_diagrams}.
By the general property of ellipses,
this is $b / a$ of the lightly shaded region bounded by the auxiliary circle,
where $b$ is the semi-minor axis.
That lightly shaded area is
the sector with area $\tfrac{1}{2} a^2 E$ minus
the triangle of base $ea$ with area $\tfrac{1}{2} e a^2 \sin(E)$:
\begin{align}
  \phi &= \cos^{-1} \left[
    \dfrac{1}{e} \left(1 - \dfrac{a (1 - e^2)}{R_{\rm Roche}}\right)\right] \nonumber\\
  E &= 2 \tan^{-1} \left[
    \sqrt{\dfrac{1 - e}{1 + e}} \tan\left(\dfrac{\pi - \phi}{2}\right) \right] \nonumber\\
  t_{\rm Roche} &= \dfrac{\tfrac{1}{2} a b \left(
    E - e \sin\left(E\right)\right)}{\tfrac{1}{2} \pi a b}\, T \nonumber\\
  &= \dfrac{E - e \sin\left(E\right)}{\pi}\, T \,,
  \label{eqn:t_Roche}
\end{align}
where $T$ is the orbit period.

\begin{figure}[t]
	\centering
  \includegraphics[
    width=\columnwidth, trim={8mm 8mm 6mm 10mm}, clip]{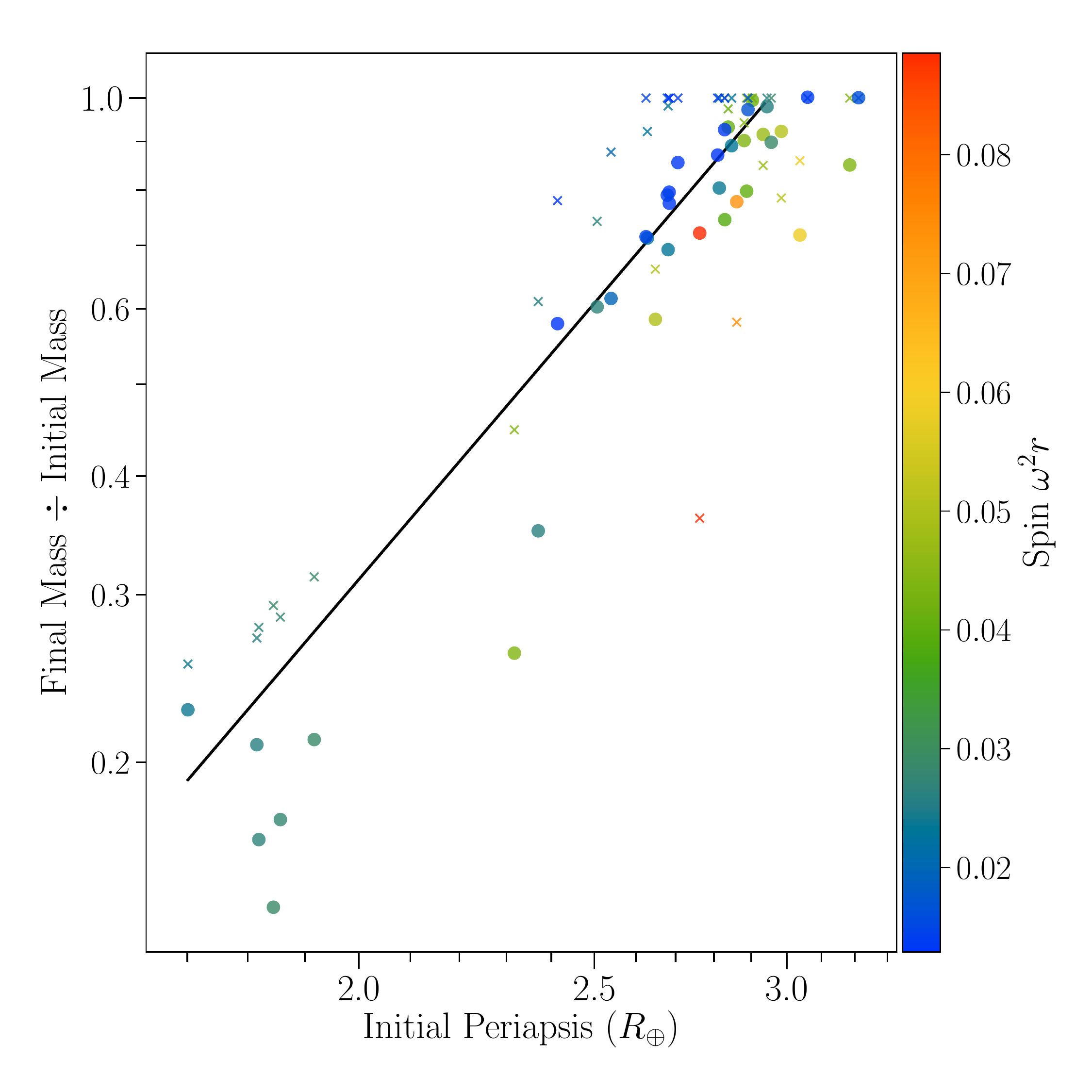}
  \caption{
    The fractional change in mass of satellites that
    are partially stripped through periapsis.
    The black line is the predicted power-law of
    $m_{\rm f}/m_{\rm i} \propto r_{\rm p}^3$,
    scaled in magnitude to fit the slow-spinning initial satellites.
    The crosses show the model predictions
    for each simulation using Eqns.~\ref{eqn:periapsis_dm},~\ref{eqn:t_Roche}.
		\label{fig:periapsis_passings_dm_r_p_spin}}
\end{figure}

The mass fractions that survive stripping from the simulations
are shown in Fig.~\ref{fig:periapsis_passings_dm_r_p_spin}
(see also Fig.~\ref{fig:periapsis_passings_m_r_p_e}).
The overall trend matches
the predicted scaling of the final mass with the cube of the initial periapsis.
Furthermore, given the simplicity of the model assumptions
compared with the dynamic complexity of the full simulated events,
the individual estimates for each initial satellite
agree remarkably well in most cases.

The vertical scatter away from the power-law trend
correlates mildly with the spin,
shown by the colour in Fig.~\ref{fig:periapsis_passings_dm_r_p_spin}.
This hints at an intuitive link
whereby rapidly rotating and more oblate satellites lose mass more readily.
However, given that the effect is relatively minor,
confirming this would warrant a study
in which the parameters are systematically varied in isolation,
since many of the different-spin cases seen here also have
different masses and other potentially degenerate properties.

\section{Exploring parameter space} \label{sec:appx:explore}

Here we discuss the effects of changing different scenario parameters
in a little more detail than in the main text.
We start with the finely sampled set of impact angles and speeds,
then explore the sets of coarser, more dramatic changes
to the initial spins of each planet,
their masses,
and their temperatures,
as detailed in \S\ref{sec:methods}.

\subsection{Changing angle and speed}

At lower impact angles (see Fig.~\ref{fig:impact_scenario})
than the examples described above, Theia is more disrupted
as it collides deeper into the proto-Earth and is kept closer in.
Plenty of material is still ejected to form an outer satellite,
but the inner remnant is too distant from the satellite
to torque it onto a sufficiently wide orbit.
The outer body may then re-impact the planet following the inner one,
resulting in a relatively smooth disk of debris,
or will graze past it and be heavily tidally disrupted into a long chain of
small, mostly unbound bodies.

At higher impact angles, Theia ploughs through the smaller mass of
the proto-Earth that it encounters with less disruption.
An outer satellite starts to form,
but it remains too close to the inner one to stay separate,
and they recombine before falling back to the planet.
This large single body can be on a wide enough orbit that it barely
recollides with the proto-Earth.
In some cases that second impact can itself produce a variety of satellites
in a similar manner to other scenarios,
including on stable orbits, but usually with quite small masses.

Changing the impact speed has a less dramatic effect.
As the speed increases, for all angles,
the overall behaviour is similar,
with both the inner remnant and outer satellite
forming farther from the proto-Earth and from each other.
This extra separation can weaken the torque between them,
but it conversely allows significantly more time for interaction,
so tends to place the satellites on closer to stable orbits.
The general result for faster impacts is thus
that similar-mass satellites form with mildly different trajectories.
Large inner remnants that would re-impact the proto-Earth
may instead pass by on a near-miss
and be dispersed into many small bodies in a large tidal tail.
Outer satellites that would be highly disrupted by tides
may reach a survivable or even Roche-exterior orbit.

At significantly higher speeds,
the sprayed ejecta become unbound and/or too dispersed to form cohesive bodies,
and the Theia remnant itself eventually escapes as a hit-and-run.
There is a middle ground where a mostly intact impactor
remains bound on a large orbit to return later,
which effectively becomes a new impact of its own
with much the same range of possible outcomes \citep{Asphaug+2021}.

\subsection{Changing pre-impact spin}

The pre-impact rotation of both the target and impactor
can have a significant effect on the outcome
of giant impacts in general
\citep{Canup2008,RuizBonilla+2021,Asphaug+2021},
especially for the relatively rapid spins in these simulations.
Regardless, we find that spinning bodies still produce
stable immediate satellites in similar ways,
and furthermore open up wider regions of the parameter space
than non-rotating ones.
We denote spins with a positive angular momentum vector
in the same direction as the orbital angular momentum as \emph{prograde},
and opposite, negative ones as \emph{retrograde}.

For a prograde proto-Earth and a given impact angle,
the overall behaviour is similar to
a higher-angle non-spinning scenario.
Theia collides with mantle material that is moving in the same direction,
so is less disrupted,
and a satellite forms slightly later on in time.
For both the $+0.25$ and $+0.5~L^{\rm max}$ targets,
all tested angles $\leq$$45^\circ$ and $\leq$$46^\circ$ respectively
produce bound satellites outside the Roche limit.
The total angular momentum is increased above
the canonical $\sim$1.2~$L_{\rm EM}$ to $\sim$2.7 and $\sim$1.9 respectively,
comparable to other high angular momentum impact scenarios
\citep{Cuk+Stewart2012,Canup2012,Lock+2018},
though any intermediate values should also be viable in this case.

Conversely, a retrograde proto-Earth greatly disrupts the impactor
into a spray of debris.
This can collect into one or two clumps,
but these tend to be too separated to fall into wide enough orbits as easily.
Some scenarios do still produce satellites on stable but eccentric orbits.
Furthermore, the total angular momentum is around half the present-day value,
so a rapidly retrograde-spinning proto-Earth is not a viable scenario.

A prograde Theia can somewhat `roll' over the proto-Earth,
but otherwise the behaviour is similar to
the non-spinning cases as the angle changes.
As this spin increases further, the outer satellite tends to be ejected unbound.
Theia's small mass means even a rapid spin has only a small contribution
to the total angular momentum.

A retrograde Theia sticks closer to the proto-Earth
after the initial impact,
but can also be disrupted enough for a separate outer body to form successfully.
For $L = -0.25~L^{\rm max}$, the torque from the nearby inner remnant
is strong enough to produce a stable satellite across a wider range of angles
than the non-spinning case.

We did not explore the great variety of other options for the initial spins,
most notably having both planets spin
or allowing spin angular momenta not parallel to the orbital angular momentum
beyond the proof-of-concept example shown in Fig.~\ref{fig:L075y_snap}.
For now, we simply conclude that pre-impact spin has a fairly significant effect
and can increase the viable region of parameter space
for the immediate formation of stable satellites.

\subsection{Changing impactor mass}

For a fixed impact angle and speed,
reducing Theia's mass so significantly relative to the proto-Earth
produces very different outcomes.
However, at larger impact angles,
much of the same fiducial-mass behaviour is then reproduced.
The range of ideal satellite-producing angles is slightly narrower,
but for the $\tfrac{3}{4}$ mass Theia
we find the same trends for the formation
of separate inner and outer bodies as the impact angle and speed change,
including the placement of large satellites
onto stable orbits well beyond the Roche limit.
At these higher angles, the satellite and inner remnant tend to form
further from the proto-Earth and from each other,
so end up on more eccentric initial orbits.

The $\tfrac{1}{2}$ mass Theia also shows similar qualitative trends,
at correspondingly even higher angles.
However, the outer body then forms too far behind the inner remnant,
and is torqued onto an eccentric orbit
without the periapsis being raised outside the Roche limit.
It might be that a narrower range of angles and speeds exists
where a stable satellite is produced,
but at best the likelihood decreases somewhat for
an impactor with a mass significantly lower than the canonical Theia's
by more than a few tens of percent.

\subsection{Changing temperature}

We probe the sensitivity to the internal structure of the two planets
by significantly lowering and raising
the surface temperatures of both planets, here by $\pm 1000$~K.
This corresponds to increasing and decreasing their densities
by about 10 and 4\%, respectively.
Given the uncertainty in this aspect of the initial conditions,
it is encouraging that
the lower (higher) temperature simulations
generally show similar outcomes,
with a mild trend of the inner remnants staying
closer to (further from) the planet.
This is as might be expected from the relative ease with which
warmer and lower density planets
can push through the initial impact.
This results in the higher temperature satellites being torqued
slightly more readily onto wider orbits.

All of these simulations use the same ANEOS materials.
However, similar satellite-forming results were found with previous simulations
using the simpler \citet{Tillotson1962} equation of state
and an even lower surface temperature of 500~K \citep{RuizBonilla+2021},
further suggesting that our conclusions are not highly sensitive to the
composition and density or thermal profiles of the planets.

\section{Data tables} \label{sec:appx:tables}

Table~\ref{tab:results} lists the primary results from all simulations,
arranged in groups by subset, angle, and speed.
Full simulation data will be shared on reasonable request.
Note that the satellite values refer to
only the single most massive body, if any;
some scenarios also produce smaller satellites on stable orbits,
especially if the largest one tabulated here is escaping unbound.
Note also that the qualitative evolution of some scenarios
has not progressed as far as others.
For example, if the initial impactor remnant is set
on an orbit with a period greater than 50 hours,
then it will not yet have re-impacted the target.
Similarly, long-period satellites
may be headed towards significant future stripping events at periapsis.

Table \ref{tab:results} is published online in its entirety in the
machine-readable format. A description of the fields in the online table
is shown here for guidance regarding its form and content.

\begin{table*}[t]
  \begin{center} \begin{tabular}{clll}
		\hline
		\hline
    Column & Parameter & Units & Description \\
		\hline
      1 &  Type                       &                      & changed parameters from the base scenario$^*$ \\
      2 &  $\beta$                    &  $^\circ$            & impact angle \\
      3 &  $v_{\rm c}$                &  $v_{\rm esc}$       & speed at contact \\
      4 &  $m$                        &  $M_\moon\!$         & mass of the largest satellite \\
      5 &  $r_{\rm p}$                &  $R_\oplus$          & periapsis of the largest satellite \\
      6 &  $e$                        &                      & orbital eccentricity of the largest satellite \\
      7 &  $f_{\rm c}$                &  \%                  & mass-fraction of core iron material of the largest satellite \\
      8 &  $m_{\rm d}^{a_{\rm eq}}$   &  $M_\moon\!\!\!$     & mass of the debris disk using the equivalent circular radius criterion \\
      9 &  $m_{\rm d}^{r_{\rm p}}$    &  $M_\moon\!\!\!$     & mass of the debris disk using the periapsis criterion \\
     10 &  $L_{\rm bnd}$              &  $L_{\rm EM}\!\!$    & total angular momentum of all bound material \\
     11 &  $f_{\rm t}^{\rm s}$        &  $\%$                & mass-fraction of target proto-Earth mantle ($f_{\rm t}$) of the full satellite \\
     12 &  $f_{\rm t}^{\rm s,70}$     &  $\%$                & $f_{\rm t}$ of the outer regions of the satellite above 70\% of its radius (roughly $\tfrac{2}{3}$ by mass) \\
     13 &  $f_{\rm t}^{\rm s,90}$     &  $\%$                & $f_{\rm t}$ of the outer regions of the satellite above 90\% of its radius (roughly $\tfrac{1}{4}$ by mass) \\
     14 &  $f_{\rm t}^{\rm p}$        &  $\%$                & $f_{\rm t}$ of the full planet \\
     15 &  $f_{\rm t}^{\rm p,85}$     &  $\%$                & $f_{\rm t}$ of the planet outside of 0.85~$R_\oplus$ by radius (roughly $\tfrac{1}{2}$ by mass) \\
     16 &  $f_{\rm t}^{\rm d}$        &  $\%$                & $f_{\rm t}$ of the debris disk \\
    \hline
	\end{tabular} \end{center}
	\caption{Results from the simulations,
    available online in machine-readable format.
    \\
    $^*$The `type' notes the other parameters
    that are changed from the base scenario
    for the following subsets of simulations,
    as described in \S\ref{sec:methods},
    within which the angle and speed are then varied, namely:
    the number of particles, $N$ (base $10^7$);
    the spin angular momentum of the target or impactor,
    $L_{\rm t, i}$ (base 0);
    the mass of the impactor, $M_{\rm i}$ (base $0.133~M_\oplus$);
    and the surface temperature, $T_{\rm s}$ (base 2000~K).
    \label{tab:results}}
\end{table*}

\end{document}